\documentclass{Interspeech2024}

\interspeechcameraready

\title{CEC: A Noisy Label Detection Method for Speaker Recognition}

\name{Yao}{Shen}
\name{Yingying}{Gao}
\name{Yaqian}{Hao}
\name{Chenguang}{Hu}
\name{Fulin}{Zhang}
\name{Junlan}{Feng$^{\ast}$}
\name{Shilei}{Zhang$^{\ast}$}

\address{
  China Mobile Research Institute, China}
\email{\{shenyao, gaoyingying, haoyaqian, huchenguang, zhangfulin, fengjunlan, zhangshilei\}@chinamobile.com}


\usepackage[ruled,linesnumbered]{algorithm2e}
\usepackage{ragged2e} 
\usepackage{booktabs,makecell, multirow, tabularx}

\begin{document}

\maketitle

\newcommand\blfootnote[1]{
\begingroup
\renewcommand\thefootnote{}\footnote{#1}
\addtocounter{footnote}{-1}
\endgroup
}
\blfootnote{*Corresponding Authors}

\begin{abstract}
    
    Noisy labels are inevitable, even in well-annotated datasets. The detection of noisy labels is of significant importance to enhance the robustness of speaker recognition models. In this paper, we propose a novel noisy label detection approach based on two new statistical metrics: \emph{Continuous Inconsistent Counting} (CIC) and \emph{Total Inconsistent Counting} (TIC). These metrics are calculated through \emph{Cross-Epoch Counting} (CEC) and correspond to the early and late stages of training, respectively. Additionally, we categorize samples based on their prediction results into three categories: \emph{inconsistent samples}, \emph{hard samples}, and \emph{easy samples}. During training, we gradually increase the difficulty of hard samples to update model parameters, preventing noisy labels from being overfitted. Compared to contrastive schemes, our approach not only achieves the best performance in speaker verification but also excels in noisy label detection. 
\end{abstract}

\section{Introduction}

Noisy labels, which refer to erroneous or abnormal labeled tags present in the training data, can lead to bias and consequently degrade the model’s performance when memorized by the model \cite{zhang2021understanding,nigam2020impact}. In fact, noisy labels are often unavoidable, especially when acquiring datasets through web scraping or crowdsourcing \cite{han2018co}. 

The rise of large models in recent years has indeed spurred the demand for large-scale training data. Training speaker recognition models with massive datasets containing millions or even tens of millions of speakers is becoming the future trend. However, obtaining large-scale training data of this magnitude through meticulous manual annotation is impractical. Consequently, handling the high proportion of noisy labels in the data becomes a crucial issue.

In recent research on noisy labels, approaches can be roughly categorized into two types \cite{song2022learning,huang2019o2u}: (a) Enhancing the robustness of the model to noisy labels through \textit{robust architectures} \cite{goldberger2016training, borgstrom2020bayesian}, \textit{robust regularization} \cite{xia2020robust, zheng2019towards}, or \textit{robust loss design} \cite{lyu2019curriculum, tong2021automatic, han2022self, li2022speaker} to reduce the model's overfitting to noisy labels. This method does not involve noisy label detection and directly trains on the unclean datasets. (b) \textit{Sample selection} is performed by utilizing the characteristic of higher loss from noisy labels. This is achieved through methods such as multi-network learning \cite{han2018co, jiang2018mentornet, yu2019does} and multi-round learning \cite{huang2019o2u, qin2022simple} to filter out potential noisy labels.

Sample selection methods, such as \textit{Co-teaching} \cite{han2018co}, \textit{Co-teaching+} \cite{yu2019does}, and \textit{O2U-Net} \cite{huang2019o2u}, offer advantages in detecting noisy labels through high loss in image classification tasks. However, their effectiveness in speaker recognition tasks remains unverified, and they rely on a relatively accurate proportion of noisy labels provided before training, which poses challenges in practical scenarios. The OR-Gate \cite{fang23_interspeech} utilizes a top-k mechanism for sample selection in speaker recognition, but its performance on complex data augmentation datasets requires validation. Furthermore, these methods filter out noisy labels in the late stages or after training, potentially allowing the model to fit some noisy labels before detection.

The factors mentioned above motivated us to undertake the work presented in this paper. The main contributions of this paper are as follows:

\begin{itemize}

\item We propose a novel noisy label detection method called \textit{Cross-Epoch Counting} (CEC). During training, samples are categorized into three groups based on model predictions: inconsistent, hard, and easy. The number of times a sample is classified as an inconsistent sample across epochs is counted by two metrics: \textit{Continuous Inconsistent Counting} (CIC) and \textit{Total Inconsistent Counting} (TIC). CIC is a continuous count, which restarts counting when samples are classified as hard or easy samples, mainly screening out samples with a high probability of being noisy labels before they are fitted by the model in the early stages of training. TIC, on the other hand, is an accumulated count that does not reset due to classification fluctuations and screens out noisy labels that CIC missed in the later stages of training. This approach does not require additional network, round, or data information and is simple yet effective.

\item During the training process, the difficulty of hard samples used for updating model parameters is gradually increased. This can effectively prevent noisy labels in hard samples from entering gradient backpropagation in the early stages of training.

\item We tested and compared several commonly used noisy label detection methods in the field of computer vision on the speaker recognition dataset.

\end{itemize}

%##################################################################

\section{Methods}

\begin{figure*}[t]
  \centering
  \includegraphics[width=\linewidth]{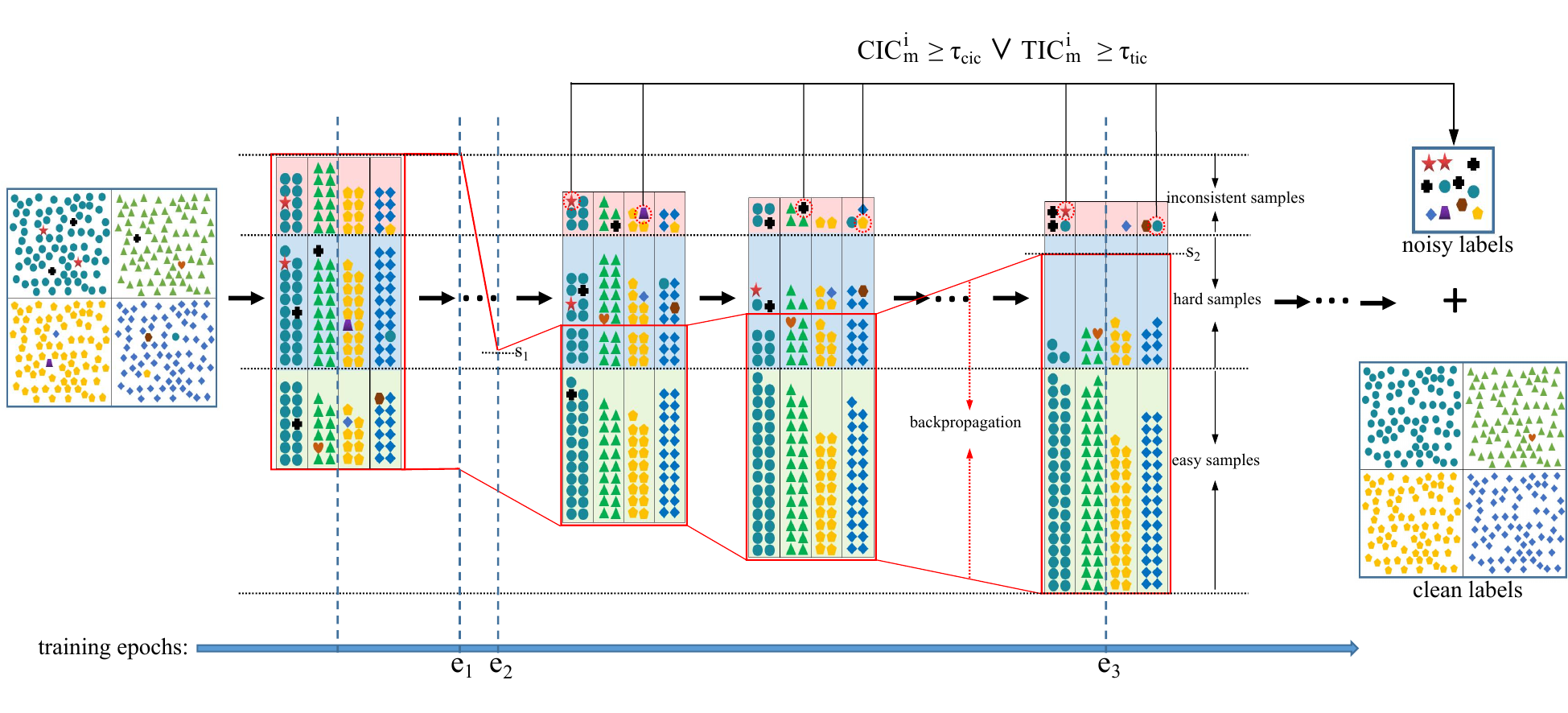}
  \caption{\textbf{Framework of the proposed method.} Samples are classified into three categories: \textbf{inconsistent}, \textbf{hard}, and \textbf{easy}. The difficulty of hard samples for gradient backpropagation is adjusted by tuning the retention threshold \(\tau_m\). From epoch \(e_2\) to \(e_3\), \(\tau_m\) gradually increases from \(s_1\) to \(s_2\), gradually increasing the difficulty of hard samples until most of them are retained for gradient backpropagation. Easy samples always participate in gradient updates, while inconsistent samples only participate in gradient updates before epoch \(e_1\). Samples classified as inconsistent are subjected to \textbf{Continuous Inconsistent Counting (CIC)} and \textbf{Total Inconsistent Counting (TIC)}. If either of these counts exceeds its corresponding threshold, the sample is considered a noisy label and immediately removed from the training samples.}
  \label{fig:fig01}
\end{figure*}

The CEC method we proposed aims to accurately detect noisy labels while removing them as early as possible. 
In each epoch, samples are classified into three categories: inconsistent, hard, and easy. The number of times a sample is classified as an inconsistent sample is counted in each epoch using CIC and TIC, and noisy labels are filtered out based on these counts. Through predefined curriculum learning, the difficulty of hard samples for updating model parameters is gradually increased during training. The whole framework is shown in Figure~\ref{fig:fig01}.

\subsection{Samples classification}

Each epoch of training will classify training samples into three categories based on the model's predictions: inconsistent samples, hard samples, and easy samples. The classification method is shown in Equation~\ref{eq:eq01}.

\begin{equation}\label{eq:eq01}
C_i = 
    \begin{cases}
        2, & if \  y_i \neq \Tilde{y_i} \\    
        1, & if \  y_i = \Tilde{y_i}, \ s_{P} < \tau_{P} \bigvee 
         s_{N} > \tau_{N} \\
        0, & otherwise.  
    \end{cases}
\end{equation}
where \(C_i\) represents the sample classification of the i-th sample, \(y_i\) is the label for this sample, \(\Tilde{y_i}\) is the predicted label by the model, \(s_{P}\) represents the cosine distance between the feature vector of the sample and the weight vector of the labeled speaker, \(s_{N}\) is the maximum cosine distance between the feature vector of the sample and the weight vectors of other speakers. \(s_{P}\) and \(s_{N}\) are obtained during the calculation of the loss. \(\tau_{P}\) and \(\tau_{N}\) are hyperparameters for thresholds, which can be determined based on the task scenario or empirical values. The values 0, 1, and 2 for \(C_i\) represent easy, hard, and inconsistent samples, respectively.

The loss function adopts AAM-softmax \cite{wang2018cosface}, as shown in Equation~\ref{eq:eq02}.

\begin{equation}\label{eq:eq02}
L = - \frac{1}{N} \sum_{i=1}^{N}{log\frac{e^{s(\cos(\theta_{y_i,i})-d)}} {e^{s( \cos(\theta_{y_i,i})-d)} +  \textstyle \sum_{j \neq y_i}{e^{s \cos(\theta_{j,i})}}}}
\end{equation}
where \(N\) is the number of training samples, \(s\) is the scaling factor, \(d\) is margin, \(\theta_{j,i}\) is the angle between \(W_j\) and the i-th feature vector and the \(W_j\) is the weight vector of the j-th class. We can consider that \(s_{P} = \cos(\theta_{y_i,i})\) and  \(s_{N} = \max  \left \{ \cos(\theta_{j,i}), j \neq {y_i} \right \}\).

\subsection{Noisy label detection}

We filter noisy labels based on two indicators: CIC and TIC. CIC refers to the number of times a sample is consecutively classified as inconsistent sample across epochs, and TIC refers to the cumulative number of times a sample is classified as inconsistent sample. The CIC and TIC of each sample are updated with each epoch of classification during training, following the update method illustrated in Equation \ref{eq:eq04} and \ref{eq:eq05}.

\begin{equation}\label{eq:eq04}
CIC_m^i = 
    \begin{cases}
        CIC_{m-1}^i + 1, & if \ C_i = 2 \\    
        0, & otherwise.  
    \end{cases}
\end{equation}

\begin{equation}\label{eq:eq05}
TIC_m^i = 
    \begin{cases}
        TIC_{m-1}^i + 1, & if \ C_i = 2 \\    
        TIC_{m-1}^i, & otherwise.  
    \end{cases}
\end{equation}
where \(m\) represents the current epoch number.

As shown in Equation \ref{eq:eq06}, if either TIC or CIC exceeds its corresponding threshold, the sample will be classified as a noisy label and removed from the training set.

\begin{equation}\label{eq:eq06}
f_{NL}(x_i) = CIC_m^i > \tau_{cic} \bigvee TIC_m^i > \tau_{tic}
\end{equation}
where \(\tau_{cic}\) and \(\tau_{tic}\) are thresholds for CIC and TIC respectively, with the value of \(\tau_{cic}\) being less than \(\tau_{tic}\). The mathematical symbol \(\bigvee\) represents logical OR. \(\tau_{cic}\) usually corresponds to the number of epochs in the early stages of the model, and \(\tau_{tic}\) can typically be equal to the number of epochs in the later stages of model training. Since the model tends to prioritize fitting clean labels in the early stages of training, and only starts fitting noisy labels in the middle to later stages \cite{bai2021understanding}, noisy labels often remain classified as inconsistent samples for multiple epochs in the early stages of training. It's only in the mid to later stages when they are fitted by the model that they become hard or easy samples. Therefore, CIC can effectively filter out most noisy labels in the early stages of training with a small threshold \(\tau_{cic}\). TIC, on the other hand, uses a larger threshold \(\tau_{tic}\), to filter noisy labels by accumulating inconsistent counting, is less susceptible to random fluctuations and typically achieves higher recall rates.

\subsection{Curriculum learning}

Curriculum learning is a training strategy that mimics the cognitive learning process of humans, where the model starts learning from easy samples and gradually increases difficulty. This method effectively reduces the model's fitting to noisy labels. We apply curriculum learning to hard samples using a predefined retention threshold \(\tau_m\) as detailed in Equation \ref{eq:eq03}.

\begin{equation}\label{eq:eq03}
\tau_m = 
    \begin{cases}
        0, & if \ m \le e_1 \\ 
        s_1 *\frac{m-e_1}{e_2 - e_1}, & if \ e_1 < m \le e_2 \\ 
        s_1 + (s_2 - s_1) *\frac{m-e_2}{e_3 - e_2}, & if \ e_2 < m \le e_3 \\
        s_2, & otherwise.  
    \end{cases}
\end{equation}
where \(e_1\) is the epoch for warm-up, and \(e_2\) is the epoch for gradually reducing inconsistent/hard samples for the gradient backpropagation. \(e_3\) is the epoch for linearly increasing the difficulty of hard samples. \(s_1\) and \(s_2\) are the lower and upper bounds of \(\tau_m\), respectively.  \(m\) is the current epoch number. \(s_1\) and \(s_2\) can be determined based on the task scenario or empirical values.

We use \(1-s_P\) to denote the difficulty of the sample. Only when \(1-s_P < \tau_m\), the hard sample will be used to update the model parameters. We do not use the loss value as the difficulty of samples because the range of variation in loss values is large, while the range of variation in cosine distance is fixed and unaffected by the data and model framework. Moreover, classifying samples based on cosine distance aligns more intuitively with the actual application of speaker recognition.

%##################################################################
%###### 3 实验
\section{ Experiments}

\subsection{Datasets and Experiments setup}

\subsubsection{Datasets}

We tested the performance of the proposed method in real-world scenarios on the VoxCeleb2 \cite{Chung18b} dataset. To evaluate the performance under different ratios of noisy labels, we extracted 2000 speakers from the VoxCeleb2 dataset, comprising a total of 162,901 utterances, as clean labels dataset. The remaining speakers were considered as open-set noisy labels, and we simulated the pair noise by randomly adding them to clean dataset according to different Noisy-to-Clean Ratio (NCR) values ranging from 0\% to 50\%. We used the Original, Extended, and Hard VoxCeleb1 \cite{Nagrani17} test sets as evaluation sets (Vox1-O, Vox1-E, and Vox1-H).

\subsubsection{Performance evaluation methods}

We evaluate the accuracy of the speaker recognition model using the Equal Error Rate (EER). F1-score, Accuracy, Precision, and Recall are used as evaluation metrics for assessing the effectiveness of noisy label detection.

%3.1.3 实验参数等
\subsubsection{Experiments setup}

In this paper, both the training code and baseline are derived from Wespeaker \cite{wang2023wespeaker}. The backbone network chosen is the ECAPA-TDNN \cite{desplanques2020ecapa} with 512 channels. The loss function employed is AAM-softmax, with a margin of 0.2 and a scale of 32. The pooling layer used is attention pooling. Features are extracted using 80-dimensional Fbank, and the embedding dimension is set to 192. The MUSAN \cite{snyder2015musan} and RIR \cite{ko2017study} datasets are respectively used for adding background noise and reverberation in on-the-fly data augmentation. Audios are randomly chunked. All training runs for 150 epochs.

According to the requirements of the actual speaker recognition application scenario for cosine distance, we set \(\tau_P = 0.6\), \(\tau_N = 0.4\), \(s_1 = 0.6\) and \(s_2 = 1.0\). \(e_1\) is the warm-up stages, and \(e_2\) can be 2 to 5 epochs larger than \(e_1\). We set \(e_1 = 6\), \(e_2 = 10\) and \(e_3 = 100\). In our test datasets, the first 20-30 epochs represent the early training stages, where accuracy rapidly improves, while the 90-100 epochs represent the mid-training stages, where accuracy gradually stabilizes. Therefore, we set \(\tau_{cic} = 25\) and \(\tau_{tic} = 95\) accordingly.

\subsection{Experimental results}

\subsubsection{Speaker recognition performance}

\begin{table}[ht]
    \caption{The comparison of EER (\%) results for speaker recognition performance on the VoxCeleb2 dataset.}
    \centering
    \begin{tabular}{cccc}
        \toprule
        \textbf{Methods} & \textbf{vox1-O} & \textbf{vox1-E} & \textbf{vox1-H} \\
        \midrule
        Baseline & 1.069 & 1.209 & 2.310 \\
        Co-teaching & 1.016 & 1.239 & 2.275 \\
        Co-teaching+ & 1.180 & 1.426 & 2.600 \\
        O2U-Net & 1.067 & 1.341 & 2.457 \\
        OR-Gate & 1.080 & 1.256 & 2.346 \\
        CEC & \textbf{1.010} & \textbf{1.199} & \textbf{2.233} \\
        \bottomrule
    \end{tabular}
    \label{tab:my_label}
\end{table}

As shown in Table \ref{tab:my_label}, Co-teaching and our algorithm achieved good performance in speaker recognition on the VoxCeleb2 dataset. However, since VoxCeleb2 itself contains only a small number of noisy labels, the improvement for both methods is not significant. Qin et al. \cite{qin2022simple} obtained that VoxCeleb2 contains about 2.6\% noisy labels through clustering, which is consistent with our results. The poor performance of Co-teaching+ may be attributed to its ``Update by Disagreement" approach, where a large proportion of disagreements between the two networks are noisy labels. OR-Gate also did not achieve satisfactory results, possibly because the optimal hyperparameters mentioned in the paper were determined without extensive data augmentation. O2U-Net did not demonstrate its advantage because it requires multi-round of training, while all our training was limited to 150 epochs, which is not sufficient for O2U-Net.

\begin{figure}[ht]
  \centering
  \includegraphics[width=\linewidth]{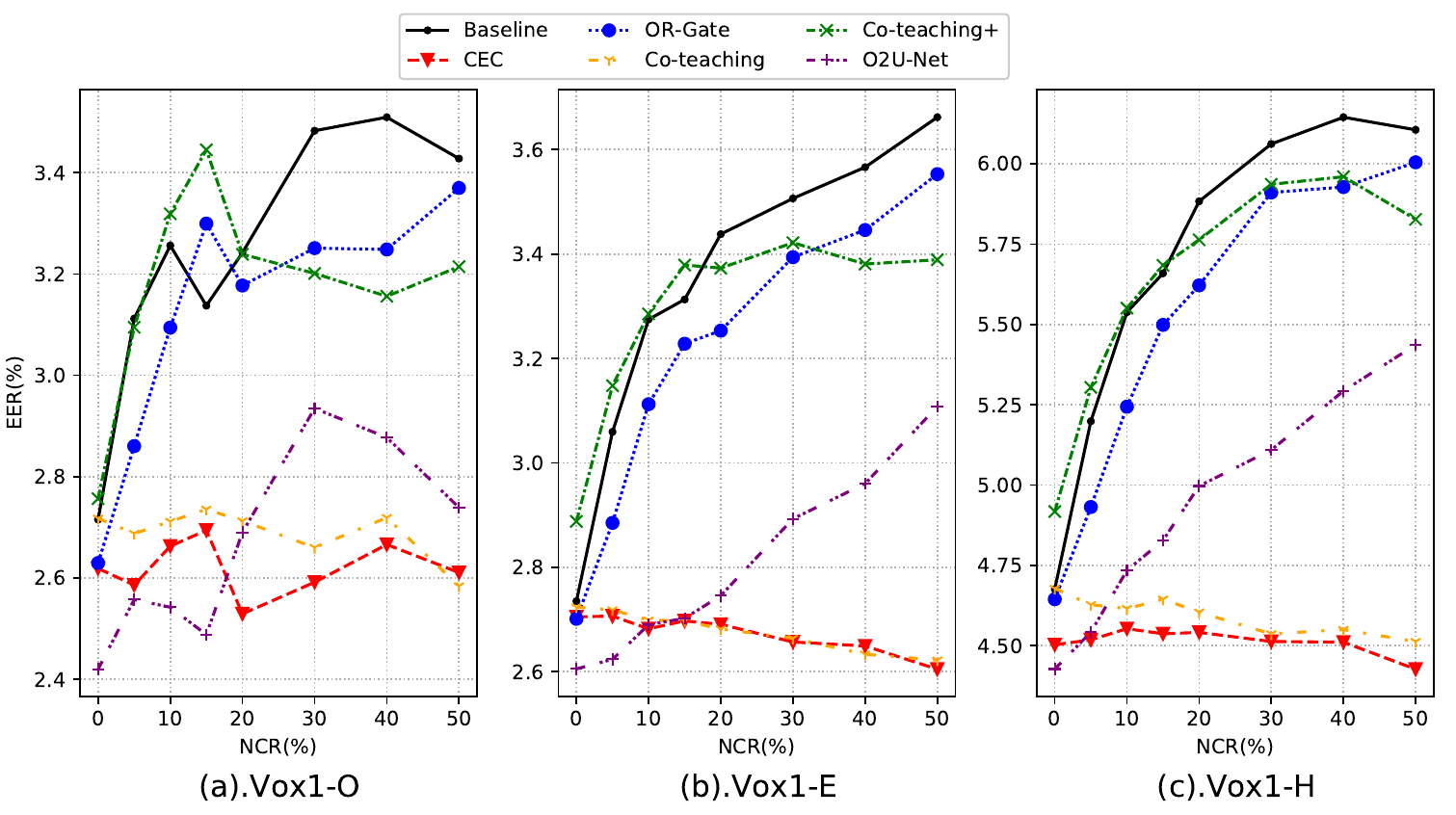}
  \caption{The comparison of EER (\%) results for speaker recognition performance on datasets with different NCRs.}
  \label{fig:fig05}
\end{figure}

The results on datasets with different NCRs are shown in Figure \ref{fig:fig05}. It can be observed that both Co-teaching and our algorithm demonstrate good robustness to increasing NCR values, and in comparison, our approach performs better on the Vox1-O and Vox1-H test sets. O2U-Net performs well only on relatively clean datasets. Co-teaching+ and OR-Gate exhibit similar issues on the VoxCeleb2 dataset, resulting in suboptimal performance.

It is important to note that in other studies, the noisy label ratio \(p\) is typically increased gradually while keeping the total data volume constant. This results in a relative decrease in clean labels, making it difficult to determine whether the deterioration in model performance is due to an increase in noisy labels or a decrease in clean labels. In contrast, we increase the number of noisy labels while keeping the number of clean labels constant, which enables us to better reflect the algorithm's robustness to different noisy label ratios. This is also why we use NCR instead of \(p\). We can consider that \(p=\frac{NCR}{NCR+1}\).

% 3.2.1
\subsubsection{Noisy label detection performance}

\begin{figure}[ht]
  \centering
  \includegraphics[width=\linewidth]{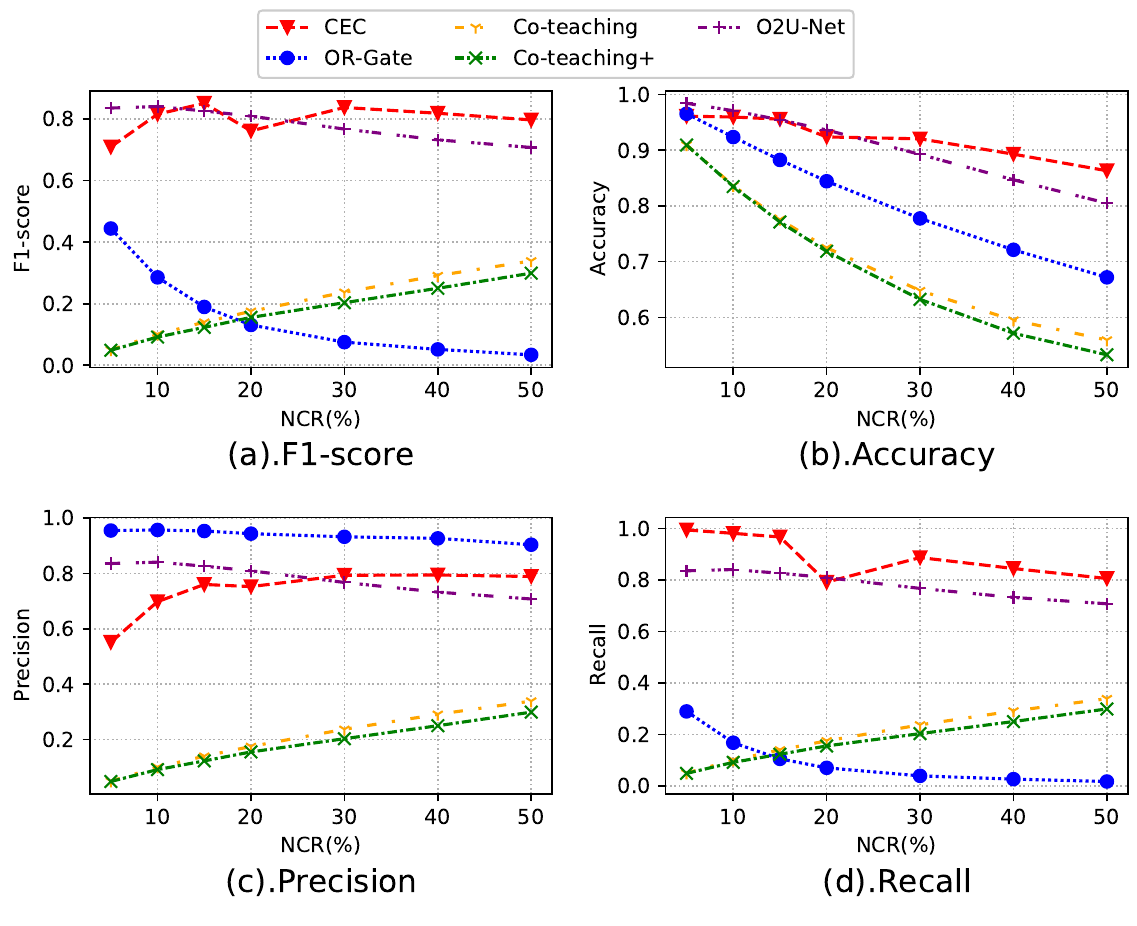}
  \caption{The comparison of results for noisy label detection performance on datasets with different NCRs.}
  \label{fig:fig06}
\end{figure}

The noisy label detection results of each algorithm under different NCRs are shown in Figure \ref{fig:fig06}. O2U-Net and our algorithm perform well. In comparison, O2U-Net performs better at low NCRs, while our method performs better at higher NCRs. This is because some clean labels may be misclassified as noisy labels by our algorithm. When NCR is low, the proportion of such data is relatively high, which distorts the results of F1-score, Accuracy and Precision, while Recall remains unaffected. Through manual review, we found that these samples include abnormal audio such as crying, laughing, screaming, whispering, etc. Although these utterances are not noisy labels, in small datasets lacking diversity, the model finds it difficult to fit these challenging samples, thus misclassifying them as noisy labels. O2U-Net effectively reduces false positives and avoids this issue when provided with the correct noisy label proportions.

Co-teaching performs well in terms of speaker recognition performance, but its ability to detect noisy labels is inferior to O2U-Net and CEC. This may be because Co-teaching and Co-teaching+ are based on the loss ordering results of a single epoch and filters high-loss samples according to the given proportion of noisy labels, while O2U-Net and CEC determine noisy labels based on multiple epochs. The existing parameters of OR-Gate are not suitable for multiple data augmentation scenarios, resulting in a significant number of missed noisy labels.

\subsection{Ablation experiments}

\begin{table}[ht]
    \caption{The results of the ablation experiments on the simulated dataset with an NCR of 5\%.}
    \label{tab:ablation}
    \centering
    \begin{tabular}{ccccc}
        \toprule
        \multirow{2}{*}{\textbf{Methods}} & \multicolumn{3}{c}{\textbf{EER(\%)}} & \multirow{2}{*}{\textbf{Recall}} \\
        \cline{2-4}
         ~ & \textbf{vox1-O} & \textbf{vox1-E} & \textbf{vox1-H} & ~ \\
        \midrule
        Baseline & 3.112 & 3.060 & 5.198 & - \\
        w/o CIC & 2.619 & 2.683 & 4.521 & 0.934 \\
        w/o TIC & 2.597 & 2.685 & 4.538 & 0.992 \\
        w/o CIC/TIC & \textbf{2.532} & 2.714 & 4.552 & -\\
        w/o CL & 2.763 & 2.767 & 4.650 & 0.936 \\
        % w/o CL & 3.421 & 3.538 & 5.612 & 0.814 \\
        CEC & 2.580 & \textbf{2.650} & \textbf{4.499} & \textbf{0.995} \\
        \bottomrule
    \end{tabular}
\end{table}

We further conducted ablation experiments to validate the effectiveness of each component in the proposed approach. To simulate datasets with fewer noisy labels similar to those encountered in real-world applications, we conducted ablation experiments on a 5\% NCR dataset. Table \ref{tab:ablation} displays the experimental results. The baseline represents not using any optimization schemes, while CL represents curriculum learning. Our proposed method shows an EER improvement of over 13\% compared to the baseline. Among them, curriculum learning has a greater impact on the model's speaker recognition performance, while CIC has a greater impact on noisy label detection performance. CIC can screen out more than 90\% of noisy labels in the early stages of training, while TIC has better recall performance (over 99\%), enabling it to screen out noisy labels missed by CIC in the later stages of training. Recall is used here because it more accurately reflects the performance of noisy label detection on low NCR data, as mentioned before, Recall is not affected by the presence of outlier data in the clean dataset.

\section{Conclusions}

In this paper, we propose CEC, a novel and effective framework for noisy label detection. During training, we classify samples into three categories: inconsistent, hard, and easy. The count of samples classified as inconsistent is tracked using the CIC and TIC metrics. Samples exceeding their respective threshold counts are considered noisy labels. CIC effectively filters out most noisy labels in the early stages of training, while TIC further removes the remaining noisy labels in the later stages. Additionally, we employ curriculum learning to gradually increase the difficulty level of updating model parameters for hard samples, preventing the model from fitting noisy labels. We conducted experiments on both synthetic datasets and the real dataset. Compared to other methods, our approach achieves optimal results in both speaker recognition performance and noisy label detection capability.

\bibliographystyle{IEEEtran}
\bibliography{mybib}

\end{document}